\documentclass{PoS}
\usepackage{graphicx}  % got figures? uncomment this

\def\gs{{_>\atop^{\sim}}}

\def\gtrsim{\mathrel{\hbox{\rlap{\hbox{\lower4pt\hbox{$\sim$}}}\hbox{$>$}}}} 

\title{The {\it EXIST} view of Super-Massive Black Holes in the 
Universe}

\ShortTitle{The {\it EXIST} view of SMBH}

\author{\speaker{Roberto Della Ceca}\\
        INAF-Osservatorio Astronomico di Brera in Milan, Milan, Italy\\
        E-mail: \email{roberto.dellaceca@brera.inaf.it}}

\author{G. Ghisellini, G. Tagliaferri, L. Foschini, G. Pareschi, 
         F. Tavecchio\\
        INAF-Osservatorio Astronomico di Brera in Merate, Merate, Italy}
\author{P. Coppi\\
        Department of Physics, Yale University, Yale, USA}
\author{J.E. Grindlay\\
        Harvard-Smithsonian Center for Astrophysics, Cambrdge, USA}
\author{M. Fiocchi, L. Natalucci, F. Panessa, P. Ubertini\\
        INAF-Istituto di Astrofisica Spaziale e Fisica Cosmica, Roma, Italy}
\abstract{
With its large collection area, broad-band energy coverage from optical/NIR (0.3 to 2.2 $\mu m$)
to soft/hard X-ray (0.1$-$600 keV), all-sky monitoring capability, and on-board follow-up, the proposed 
{\it Energetic X-ray Imaging Survey Telescope} mission ({\it EXIST}, see L. Natalucci contribution at this conference) has been designed to properly tackle the study of the AGN phenomenon and the role that SMBH play in the Universe. 
In particular {\it EXIST} {\it will carry out an unprecedented survey above 10 keV} (a factor $\sim$ 20 increase in hard X-ray sensitivity compared to current and prior X-ray missions) {\it of SMBH activity, not just in space but also in time and over a significant expanded energy range}; this strategy will overcome previous selection biases, 
will break the ``multi-wavelength" identification bottleneck and will dramatically increase the number of AGN detected above 10 keV that are amenable to detailed follow-up studies ($\sim 50000$ AGN are expected).
We discuss here on few selected AGN science topics enabled by the unique combination of {\it EXIST}'s instruments.
In particular {\it EXIST} will enable major progress in understanding: i) when and where SMBH are active in the Universe (by revealing and measuring heavily obscured accretion in the local - z$<0.5$ - Universe), ii) the physics of how SMBH accrete (by studying the broad-band X-ray spectra and variability properties of an unbiased and significant sample of AGN), and iii) the link between accretion power and jet/outflow power (by using observations of blazars).  Last but not least {\it EXIST}'s ability to find powerful, but very rare blazars, enables it to probe the appearance of the very first SMBH in the Universe allowing us to set strong constraints on the models of SMBH formation and early growth in the Universe.}

\FullConference{The Extreme sky: Sampling the Universe above 10 keV,\\ 
                 October 13-17 2009,\\ 
                 Otranto (Lecce) Italy}

\begin{document}

\section{Introduction}
Active Galactic Nuclei (AGN) emit over the entire electromagnetic spectrum and are widely believed
to be powered by accretion of matter onto a Super-Massive ($10^6$  to $10^{10}$ M$_{\odot}$) Black Hole (SMBH).
Besides being sites of ``extreme" physics, it is now clear that AGN are leading actors in the formation and evolution of galaxies and, in general, of cosmic structures in the Universe: fully unravelling the history of galaxy formation thus requires us to find when, where, and how SMBH grow and interact with their surroundings. The {\it EXIST} mission will be a major leap forward for AGN studies:  
using the 2 years survey data (see Figure 1) we should be able to accumulate a sample of {\it $\sim 50000$ AGN selected above 10 keV} increasing the present statistics (see \cite {cusumano2009} and references therein) a factor $\sim$100 (on average) for single classes of sources. The most interesting sources could be thus studied in point mode (about three years of the planned {\it EXIST} five years mission should be devoted to point observations) {\it using the three complementary 
instruments} 
({\it IRT}=Optical/Infrared Telescope,
{\it SXI}=Soft X-ray Imager
and 
{\it HET}=High Energy Telescope) 
{\it having unprecedented wavelength coverage} (from 0.3 to 2.2$\mu m$ and from 0.1 to 600 keV) {\it that can operate simultaneously}:
this will allow to obtain identifications, redshifts, and diagnostics of extreme objects.
Given the number of expected AGN and their {\it z} distribution (see Figure 1) we will have for the first time the possibility to investigate the cosmological properties of unabsorbed ($N_H<10^{22}$ cm$^{-2}$), mildly absorbed  ($10^{22}<N_H<10^{24}$ cm$^{-2}$), deeply absorbed ($N_H > 10^{24}$ cm$^{-2}$) AGN and blazars 
(those radio-loud AGN having the jet pointing toward us) simultaneously, with strong implication on the measurement of the accretion luminosity of the Universe and the history of SMBH growth
%and the AGN unification scheme.
%We will discuss below a few hot topics that could be addressed using these AGN samples 
(see also \cite {coppi2009}).  

\section{Unbiased Demographics of SMBHs}

It is now well established that the largest fraction of the AGN population is obscured by a large amount of cold matter around the Active Nuclei that does not permit a direct view to the central energy source: these obscured sources are fundamental for our understanding of the SMBH history (and their influence on galaxy formation)
as the large majority of the energy density generated by accretion of matter in the Universe seems to take place in obscured AGN (\cite {fabian1998}).
For absorbed AGN having column densities up to few times $10^{23}$ cm$^{-2}$, {\it XMM-Newton} and {\it Chandra} are producing a wealth of useful data (at E<10 keV) that can be used to evaluate their physical and cosmological properties (e.g. see \cite{yencho2009} and references therein). On the other hand, the situation is almost completely unconstrained for the sources having $N_H > 10^{24}$ cm$^{-2}$ (the so called Compton Thick AGN, CThick AGN hereafter, where the absorber is optically thick to Compton scattering) with not even two dozen "bona fide" CThick AGN found and studied so far \cite {dellaceca2008b}. So CThick AGN may make up anywhere from 10\% to 60\% of the absorbed AGN population (\cite {gilli2007}, \cite{dellaceca2008}, \cite {treister2009}, \cite{malizia2009}). 
Hard X-ray data (E$>$10 keV) are fundamental to unveil CThick AGN, and thus to derive a closer value of the accretion luminosity of the Universe, and so to constrain the history of SMBH growth. 

The {\it EXIST} mission is expected to make a major - {\it maybe unique} - improvement in this field:  at the 
2 year survey flux limit of $\sim 8\times 10^{-13}$ erg cm$^{-2}$ s$^{-1}$ (10-40 keV) {\it EXIST} should 
detect $\sim 1400$ CThick AGN having a redshift distribution as reported in Figure 1 (right panel), where it is compared with the expected redshift distribution of the Compton thin AGN ($N_H < 10^{24}$ cm$^{-2}$, CThin AGN hereafter) population.
We should be able to assemble a sample of $\sim 1100$ local (z<0.1) CThick AGN allowing us to derive, {\it with unprecedented accuracy}, their hard X-ray luminosity function (XLF) in the local Universe. The cosmological evolution of this XLF can thus be followed and investigated up to $z\sim$0.5 ($\sim 300$ CThick AGN with $0.1<$z$<0.5$) and, maybe, up to $z\sim$1.0 ($\sim 30$ CThick  AGN with  $0.5<$z$<1.0$). 
It is worth noting that the sample of CThick AGN could be efficiently selected and separated from the remaining CThin AGN population ($\sim 28000$ sources) using hardness ratios (e.g. the ratio between the source counts in the 5-20 and 20-40 keV energy range) from the data accumulated using the {\it EXIST/HET} instrument alone, while the source radio properties will be useful to discriminate CThick AGN from blazars ($\sim 19000$ sources). 

\begin{figure}
  \includegraphics[height=.25\textheight]{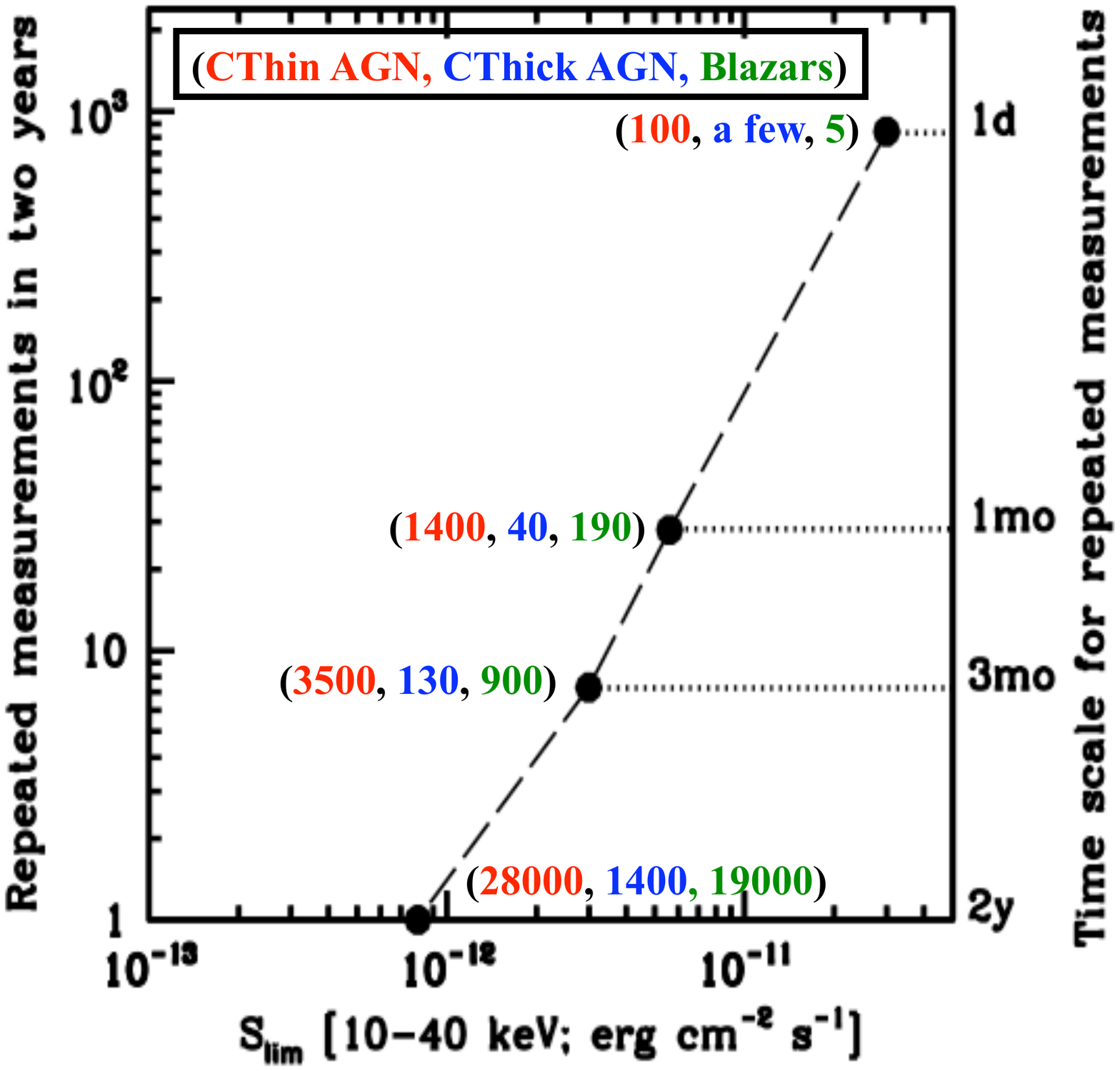}
  \includegraphics[height=.25\textheight]{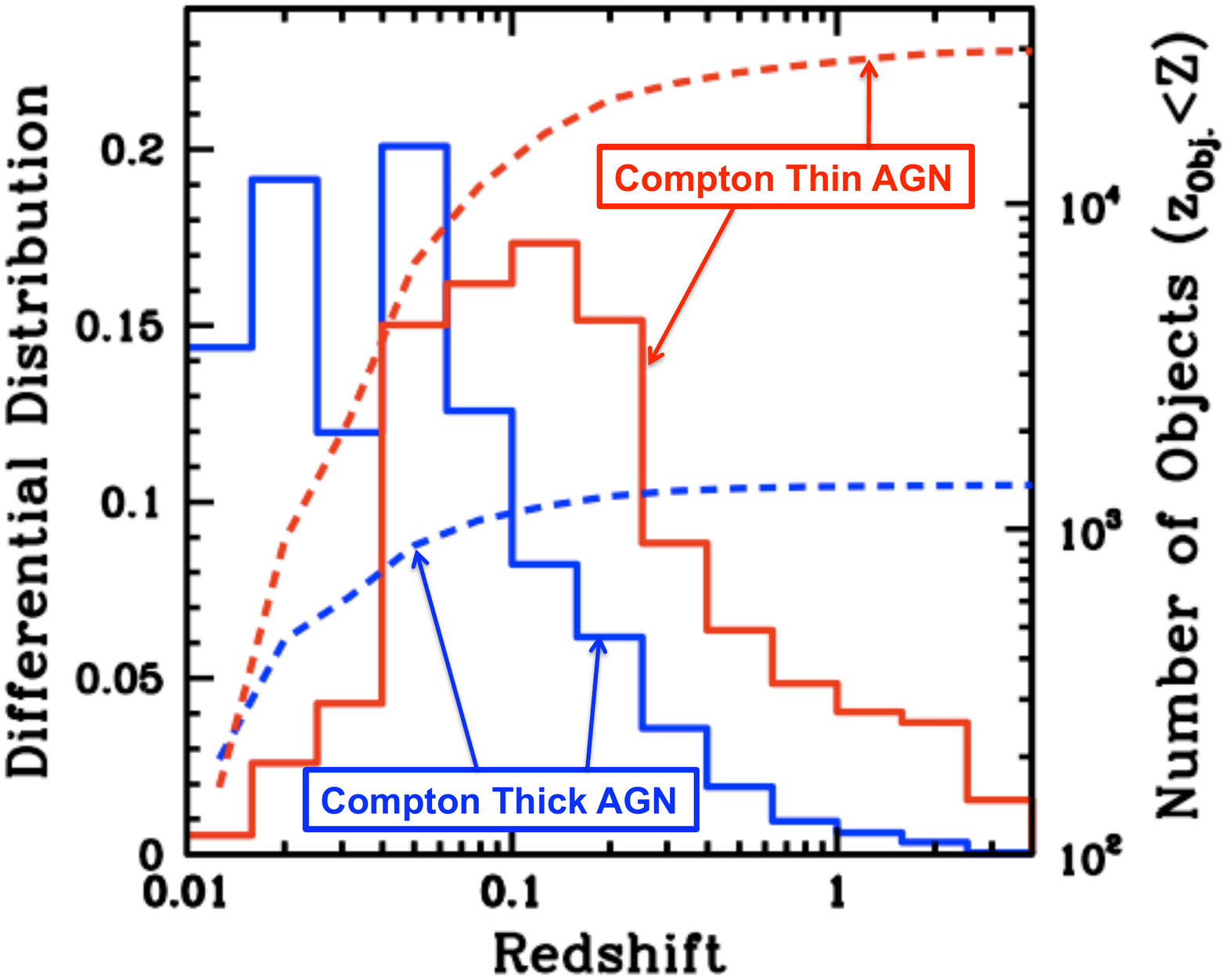}  
\vskip -0.5 truecm
  \caption{{\it Left panel:} Number of AGN that can be monitored to a given timescale using the 2 years survey data. The numbers close to the filled dots are the CThin AGN (left number), the CThick AGN (middle number) and blazars (right number) expected at that flux limit. The expected number of CThin AGN has been 
obtained by using the local XLF derived from \cite {ajello2009} in the 15-55 keV energy range ({\it Swift/BAT} data) and the cosmological evolution model described in \cite {dellaceca2008}; the expected number of CThick AGN has been derived by integrating the XLF with the cosmological evolution model discussed in \cite {dellaceca2008}, taking into account energy conversion bands, the modeling of X-ray spectra transmitted through Compton-thick absorbers and a flat $N_H$ distribution between $10^{24}$ and $10^{25}$ cm$^{-2}$; finally the number of expected blazars has been derived by extrapolating the cosmological evolution properties (XLF and evolution) of the {\it Swift/BAT} blazars as derived by \cite {ajello2009} (see \cite {ghisellini2009} for details).
{\it Right panel:} The predicted differential (solid, normalized to unity, left axis) and cumulative (dashed, right axis) redshift distributions for CThin AGN and CThick AGN at the 2 years survey flux limit (10-40 keV) of $\sim 8\times 10^{-13}$ erg cm$^{-2}$ s$^{-1}$. 
}
\end{figure}

\section{Nature, Masses \& Environment of SMBHs}

The observed AGN X-ray spectrum is complex and consists of several components, which are intimately related to the poorly understood presence of circum-nuclear matter; this matter imprints features -- emission and absorption lines, low energy cut-offs and a reflection component -- onto the primary X-ray spectrum;
X-ray observations  of a few nearby absorbed AGN have showed that this matter is highly 
structured with a range of ionization states, densities, geometries and locations (see \cite{turner2009} for a review).
Moreover the observed variability implies that the absorber has to be clumpy and at much smaller distance than 
the conventional obscuring ``torus" (see e.g. \cite{risaliti2009}). These discoveries have shown that to study source geometry, distribution, location and state of the absorbing material and intrinsic luminosity of AGN it is fundamental to disentangle all the spectral components. This in turn requires excellent broad-band coverage (from a few keV up to few hundred keV) combined with the possibility to have X-ray spectral monitoring observations, {\it such as will be provided by the {\it EXIST} mission}.
To make things worse the ``standard" AGN spectrum derived using the $\sim 50$ AGN with good broad-band X-ray coverage (see \cite{dadina2008}, \cite{molina2009}) is routinely used in important applications such as the synthesis of the Cosmic X-ray Background or to estimate AGN heating of the host galaxy ISM (and so to measure AGN feedback). 
An assessment of the X-ray spectral features of AGN based on a larger, unbiased and statistically significant sample of AGN is thus badly needed. The {\it EXIST} 
mission is {\it unique} in this respect: allowing for a 2 years survey, and making use of {\it EXIST}'s instruments for the remaining 
three years, it should be possible to accumulate good quality broad-band X-ray data for around a thousands of AGN and to derive good accuracy measurements of the AGN intrinsic spectral parameters 
(spectral slope, Compton reflection components, e-folding energy cut-off). 
Furthermore, {\it EXIST} will monitor absorption changes in about one hundred absorbed ($N_H > 10^{22}$ cm$^{-2}$) AGN 
down to a flux limit (10-40 keV)  of a few times  $10^{-11}$ erg cm$^{-2}$ s$^{-1}$, increasing the present statistics for studies of the inner structure of AGN by at least a factor 10.

The X-ray variability properties of AGN have been extensively studied during the past twenty years and significant 
progress has been achieved in the estimation of their X-ray power spectral density functions (PSD).
The AGN PSD accumulated so far ($\sim 20$ AGN) shown a $\sim -2$ power law shape at high frequencies which, 
below a characteristic ``break-frequency", $\nu_{bf}$, flattens to a slope of $\sim -1$;
moreover, AGN with known BH masses lie on the same relation between the $\nu_{bf}$ and BH mass as 
Galactic BHs, indicating some characteristic length scale proportional to BH mass (see \cite {mchardy2009} end references therein).
AGN timing and a properly calibrated $\nu_{bf}$($M_{BH}$; $L_X/L_{bol}$) relationship could provide the long sought independent black hole mass estimate that could be used to test the evolution and robustness of the 
$M_{BH}$-$\sigma$ relation at substantial redshift. Interestingly, after taking into account the 
possible variability due to absorption changes, this relation could be the only way to measure BH masses for obscured AGN. The 2-year {\it EXIST} all sky survey will give excellent quality X-ray light curve (at E$\sim 20$ keV) for $\sim 400$ AGN, spanning a BH mass range $\sim 10^7 - 10^{10}$ M$_{\odot}$, while short-term monitoring using the {\it SXI} for selected objects could drop the mass limit to $\sim 2\times 10^6$ M$_{\odot}$, i.e. {\it EXIST} can cover the bulk of AGN masses of current interest.

%----------------------------------------------------
\begin{figure}
\vskip -0.85truecm
\includegraphics[width=8.5cm]{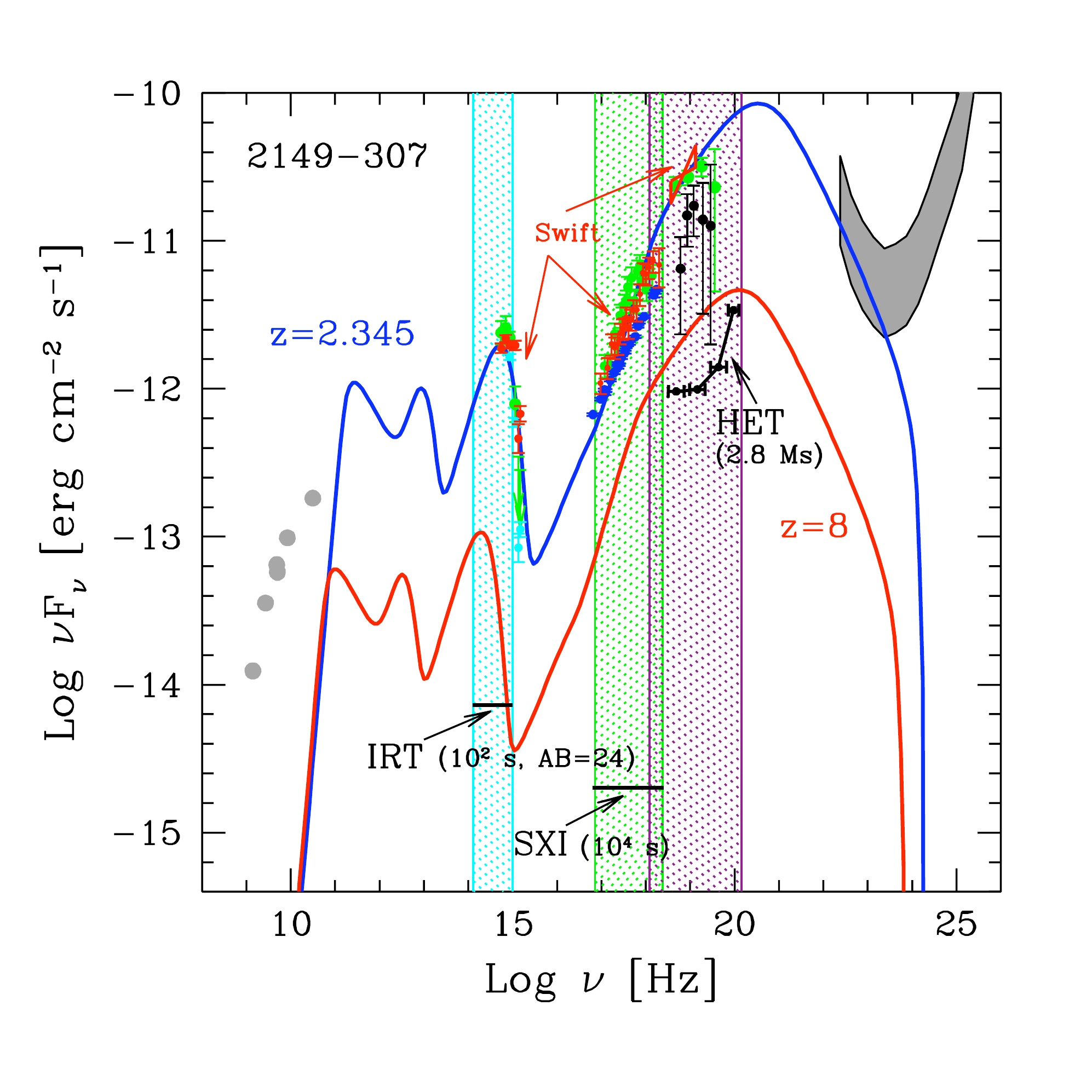}
\includegraphics[width=6.8cm]{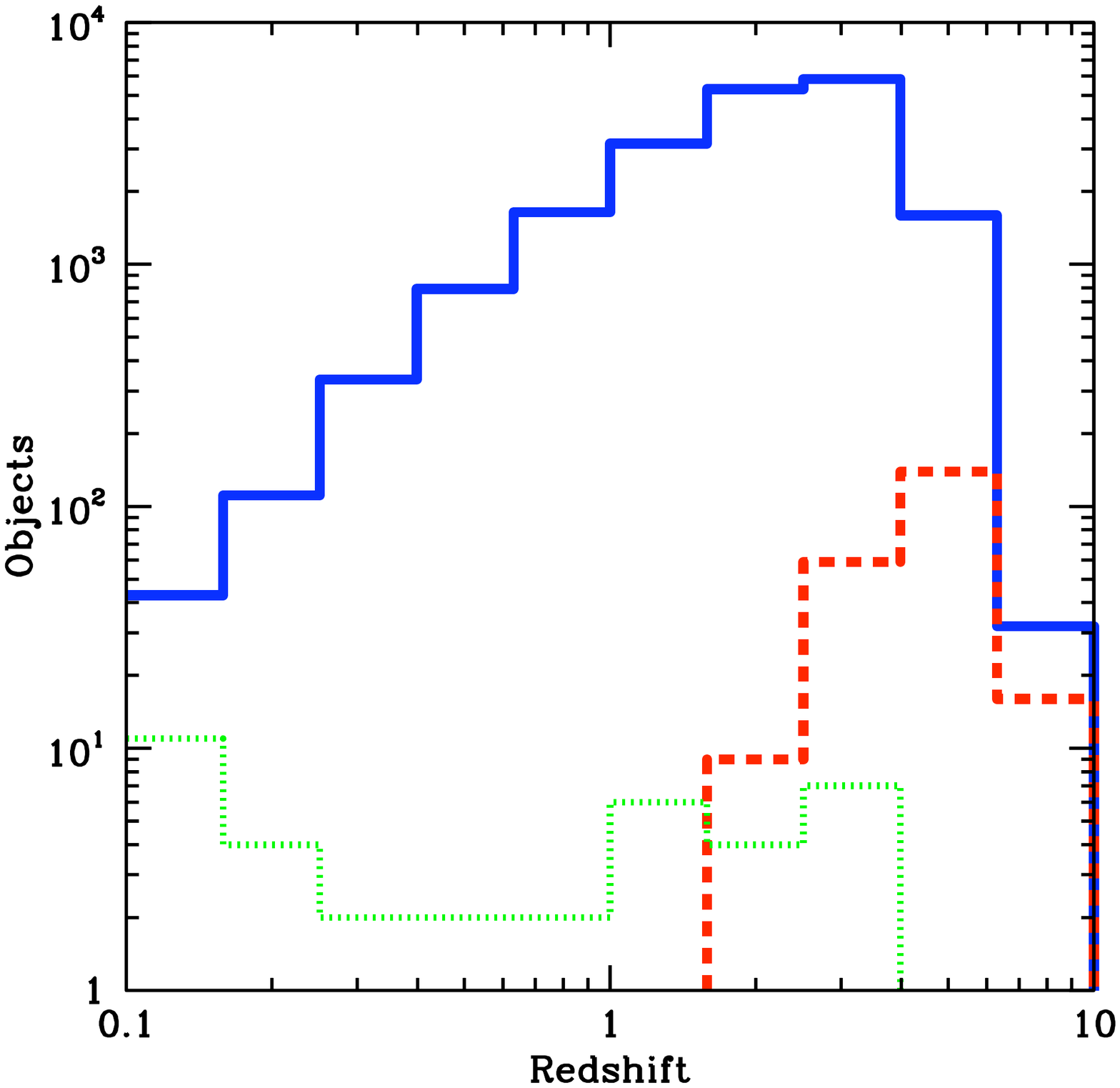}
\vskip -1.0 truecm
\caption{
{\it Left panel:}
The SEDs and model for 2149--307 at its actual redshift ($z=2.345$)
and what the model SED would appear if the source were at $z=8$.
We also show the limiting sensitivities for the three instruments foreseen
to be onboard the {\it EXIST} mission: the {\it HET} 
(5--600 keV range), the {\it SXI} (0.1--10 keV range), and the {\it IRT} 
(0.3 to 2.2 $\mu m$). We also indicate the exposure time needed to reach the shown sensitivities.
The grey stripe at high energies represents the detection limits of the {\it FERMI/LAT} instrument (lower bound: 
$5\sigma$ sensitivity limit for one year survey; upper bound: $10\sigma$ sensitivity limit for the three months survey).
It is worth noting that objects like 2149--307 at high redshift are not detectable by {\it FERMI}.
{\it Right panel} The solid line show the redshift distribution of the $\sim$ 19000 blazars detectable in the 
2 years {\it EXIST}'s all sky survey, while the dashed line show the 
redshift distribution of the high luminosity ($L_{(10-40 keV)} \gs 2\times 10^{47}$ erg s$^{-1}$, $\sim$ 220 objects expected one's . We also shown for comparison (dotted line) the redshift distribution of the 38 {\it Swift/BAT} blazars (\cite {ajello2009}).
See \cite {ghisellini2009} for details on these figures.
}
\end{figure}
%----------------------------------------------------

\section{Blazars in the {\it EXIST} era}

{\bf Jet Kinetic vs. Thermal Accretion Power through Cosmic History.}
Recent studies of extragalactic jets have shown that the power carried out by them in the form of particles and fields 
can be (on average) $\sim 10$ times larger than the luminosity radiated by the accretion process in the same
objects (\cite {ghisellini2009b}). Even if they dissipate and radiate $\sim 10\%$ of their power, relativistic effects boost their flux to such an extent to make objects point to us visible up to high redshifts. 
Moreover there are now robust indications that the jet power correlates with the mass accretion rate: this makes 
blazars (i.e. jets pointing to us) the best sources to study the relation between accretion and relativistic 
outflow. The combination of the three instruments on board {\it EXIST}, together with its ability to survey the sky, will enable {\it EXIST} to detect $\sim 19000$ blazars, even more than {\it FERMI}, that span a large range of black hole masses, accretion and jet powers, luminosities (from $10^{43}$ to $10^{48}$ erg s$^{-1}$) and redshifts (see Figure 2, right panel). According to the extrapolation of the blazars cosmological evolution model derived using {\it Swift/BAT} data (\cite {ajello2009}), more than $\sim 10000$ blazars are expected at z$>$2, $\sim 1600$ at z$>$4 and  $\sim 60$ at z$>$6 (see \cite {ghisellini2009} for details), {\it making clear the great discovery potential of {\it EXIST}}. For the first time, then, we will be able to study in a complete way how SMBH inflows and outflows (and their relation) evolve in cosmic time, a long standing and still unsolved issue, sheding light on how the jets form, collimate and accelerate.

{\bf Blazars as probes of SMBHs formation.}
It is remarkable that powerful blazars like 2149-307, if they exist even at $z=8$, could easily be detected by hard X--ray telescopes like {\it EXIST/HET} (see Figure 2, left). 
As shown in \cite {ghisellini2009} all the 10 BAT blazars with z$>$2 and $L_{(15-55 keV)}>2\times 10^{47}$ erg s$^{-1}$ have $M_{BH}>10^9\ M_{\odot}$; high X-ray luminosity blazars can be thus used as a  proxy of high mass objects with important implications on the estimate of densities of heavy black holes in the early growth of the Universe. The expected redshift distribution of high X-ray luminosity blazars 
(e.g. $L_{(10-40 keV)} \gs 2\times 10^{47}$ erg s$^{-1}$), obtained by extrapolating the blazar cosmological evolution model derived in \cite {ajello2009}, is shown in Figure 2 (right panel); if this extrapolation is correct {\it EXIST} should detect $\sim 150$ high luminosity blazars at z$>$4, 
$\sim 25$ high luminosity blazars at z$>$6, and $\sim 10$ high luminosity blazars at z$>$7.The presence of {\it SXI} and especially the IR--optical telescope would allow us to find the redshift and to provide, very easily, a complete spectrum of the accretion disk at its emission peak; this will allows to probe the appearance of the very first SMBH in the Universe,  to derive robust estimate of the accretion rate and the black hole mass and, given the survey nature of the {\it EXIST} mission, to constrain the black hole mass function of distant (z>4) radio loud AGN. 
We stress that finding even one of these sources at $z>7$  will put strong constraints on the models of SMBH formation and early growth in the Universe.

\section{{\it EXIST} Source Localization and Counterpart Identification}

Assuming a spread in the X-ray to optical (X/O) flux ratio for the AGN population from 0.1 to 10 and using the X-ray fluxes above 10 keV (where the absorption effects are less severe), the AGN found at the survey sensitivity limit are expected to have an R optical magnitude in the range between 14 and 19.5, quite bright and with the large majority of the sources {\it well within the limit of present and future spectroscopic large area/all sky surveys}. Moreover, the {\it HET} represents a major advance in coded aperture technology and can localize a 5$\sigma$ source to $\sim 20^{\prime\prime}$; so, for the large majority of the {\it HET} detected sources, the match to existing or future optical spectroscopic catalogs will 
be obvious. For the small but interesting tail in the magnitude distribution that will be significantly fainter than $R\sim 20$, e.g., powerful obscured objects at high redshifts or BL Lac objects, the optical characterization of the counterparts will be obtained using the combination of {\it SXI} and {\it IRT}. 
The bulk of the FSRQs sample (the blazars's dominant population)  
should have a radio flux greater than 0.05 Jy, so well within the limit of present radio survey (e.g. NVSS). In summary the spectroscopic identification of the {\it EXIST/HET} detected AGN will not be a major problem and, for the bulk of sources, it will be obtained ``for free" using archival data; the most interesting AGN will be then characterized in detail with dedicated {\it SXI} and {\it IRT} follow-up observations.

{\bf Acknowledgements:} We acknowledge financial support from ASI grant no. ASI I/088/06/0.

\end{document}